\documentstyle[epsfig]{aipproc}

\def\ltap{\raisebox{-.4ex}{\rlap{$\sim$}} \raisebox{.4ex}{$<$}}
\def\rts{\sqrt s}

\def\anti{\overline}

\def\mev{~{\rm MeV}}

\def\anti{\overline}

\def\mt{m_t}

\def\alt{\ltap}

\begin{document}
\hspace*{3.6in}{\bf IUHET-379}\\
\hspace*{3.6in}{\bf December 1997}
\title{The Top-Antitop Threshold\\ at Muon Colliders\thanks{Presented
at the Workshop on Physics at the First Muon Collider and at the
Front End of a Muon Collider, November 6-9, 1997,
Fermi National Accelerator Laboratory.}}

\author{M. S. Berger}
\address{Indiana University\\
Bloomington Indiana 47405}

\maketitle

\begin{abstract}
Muon colliders are expected to naturally have a small spread in 
beam energy making them an ideal place 
to study the excitation curve. We present the parameter determinations 
that are possible from measuring the total cross section near threshold at a 
$\mu ^+\mu ^-$ collider.
\end{abstract}

\section*{Introduction}

Accurate measurements of particles masses, couplings and widths are possible
by measuring production cross sections near threshold. The naturally small
beam energy spread of a muon collider would provide an excellent opportunity 
to make these measurements. Pair production of $W$-bosons, $t\bar{t}$ 
production and the Bjorken process $\mu^+\mu^-\to ZH$ have been considered as
possible places to study thresholds at a muon collider\cite{bbgh3,ttbar,Zh}.
There is very rich physics associated with the $t\bar t$ threshold, including
the determination of $m_t$, $\Gamma_t$ ($|V_{tb}|$), $\alpha_s$, and
possibly $m_h$ \cite{kuhn}.
A precise value of the top-quark mass $m_t$ could prove to be
very valuable in theoretical studies. 

\section*{Top-quark Mass Measurement at the
$\mu^+\mu^-\to \lowercase{t \bar t}$ Threshold}

Fadin and Khoze first
demonstrated that the top-quark threshold cross section is
calculable since the large
top-quark mass puts one in the perturbative regime of QCD,
and the large top-quark width effectively screens nonperturbative effects
in the final state \cite{fk}.
Such studies have since been performed by several
groups \cite{feigenbaum,kwong,sp,jht,bagliesi,sfhmn,immo,fms}.
The phenomenological potential 
is given at small distance $r$ by two-loop perturbative QCD and
for large $r$ by a fit to quarkonia spectra.
In our analysis we make use of the Wisconsin potential~\cite{wiscp} that
interpolates these regimes.

The beam energy spread 
at a $\mu^+\mu^-$ collider is expected to naturally be small. 
The rms deviation $\sigma $ in $\sqrt{s}$ is given by \cite{bbgh1,bbgh2}
\begin{equation}
\sigma = (250~{\rm MeV})\left({R\over 0.1\%}\right)\left({\sqrt s\over {\rm
350\ GeV}}\right) \;,
\end{equation}
where $R$ is the rms deviation of the Gaussian beam profile.
With $R\alt 0.1\%$ the resolution $\sigma$ is of the same
order as the measurement one hopes to make in the top mass.
For $t\overline{t}$ studies the exact shape of the beam is not
important if $R\alt 0.1\%$. We take
$R=0.1\%$ here; the results are not improved significantly with better
resolution\footnote{The most recent TESLA design
envisions a beam energy spread of $R=0.2\%$\cite{miller}, and a high 
energy $e^+e^-$ collider in the large VLHC tunnel would have a beam 
spread of $\sigma _E=0.26$~GeV\cite{norem}}.

\begin{figure}[htb]
\leavevmode
\begin{center}
\epsfxsize=3.0in\hspace{0in}\epsffile{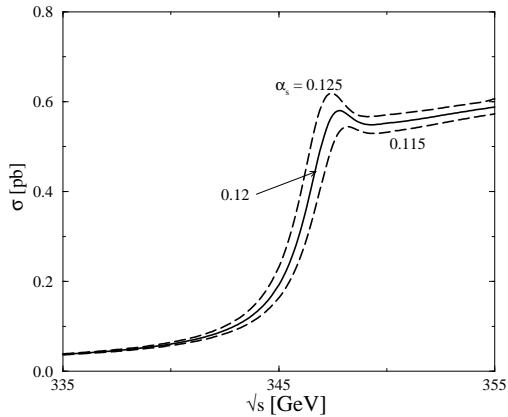}
\end{center}
\caption[]{\footnotesize\sf The cross section for $\mu^+\mu^- \to t \bar{t}$
production in the threshold region, for $m_t=175$ GeV and
$\alpha_s(M_Z)=0.12$ (solid) and 0.115, 0.125 (dashes).
Effects of ISR and beam smearing are included.}
\label{figure5}
\end{figure}

Changing the value of the strong coupling
constant $\alpha _s(M_Z)$ influences
the threshold region. Large values lead to
tighter binding and the peak shifts to lower values of $\sqrt{s}$.
Weaker coupling also smooths out the threshold peak.
These effects are illustrated in Fig.~\ref{figure5}.

\begin{figure}[htb]
\leavevmode
\begin{center}
\epsfxsize=3.0in\hspace{0in}\epsffile{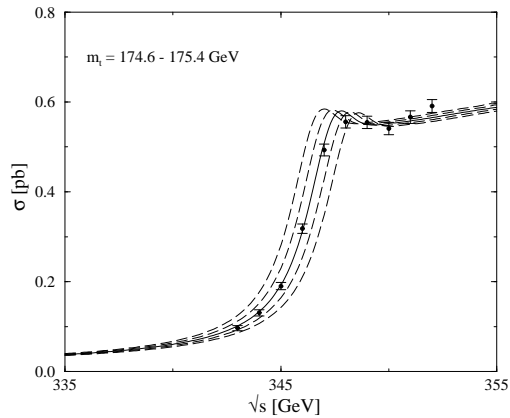}
\end{center}
\caption[]{\footnotesize\sf Sample data for $\mu^+ \mu^-
\rightarrow t \bar{t} $ obtained assuming a scan over
the threshold region devoting 10~fb$^{-1}$ luminosity to each data point.
A detection efficiency of 29\% has been assumed \cite{fms}
in obtaining
the error bars. The threshold curves correspond to shifts in $m_t$ of 200~MeV
increments. Effects of ISR and beam smearing have been included,
and the strong coupling $\alpha_s(M_Z)$ is taken to be 0.12.}
\label{figure6}
\end{figure}

\begin{figure}[htb]
\leavevmode
\begin{center}
\epsfxsize=3.0in\hspace{0in}\epsffile{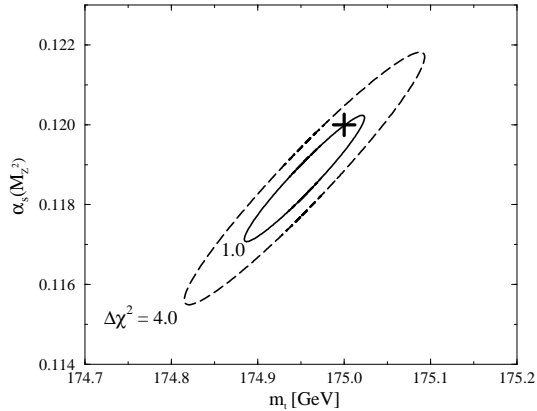}
\end{center}
\caption[]{\footnotesize\sf The $\Delta \chi ^2=1.0$ and
$\Delta \chi ^2=4.0$ confidence
limits for the sample data shown in Fig.~\ref{figure6}. The ``+''
marks the input values from which the data were generated.}
\label{figure7}
\end{figure}

To assess the precision of parameter determinations from
cross section measurements, we generate hypothetical sample data,
shown in Fig.~\ref{figure6}, assuming that 10~fb$^{-1}$ integrated luminosity
is used to measure the cross section at each energy in
1~GeV intervals. Since the top threshold curve
depends on other quantities like $\alpha_s(M_Z)$, one must do a full scan to
determine the shape of the curve and its overall normalization.
To generate the ten
data points in Fig.~\ref{figure6} we use nominal values of $m_t=175$ GeV
and $\alpha_s(M_Z)=0.12$.
Following Ref.~\cite{fms} we assume a 29\% detection efficiency for
$W\to q\bar q$, including the decay branching fraction. 
The data points can then be fit to theoretical
predictions
for different values of $m_t$ and $\alpha_s(M_Z)$; the
likelihood fit that is obtained is shown
as the $\Delta \chi ^2$ contour plot in Fig.~\ref{figure7}.
The inner and outer curves are the $\Delta \chi^2=1.0$ (68.3\%) and $4.0$
(95.4\%)
confidence levels respectively for the full 100~fb$^{-1}$ integrated
luminosity.
Projecting the $\Delta \chi^2=1.0$ ellipse on the $m_t$ axis,
the top-quark mass
can be determined to within $\Delta\mt\sim 70\mev$,
provided systematics are under control. (Systematic error issues
will be discussed later.)
A top-quark mass of 175 GeV can be measured to about 200 MeV at
$90\%$ confidence level with 10~fb$^{-1}$ luminosity.

\begin{figure}[htb]
\leavevmode
\begin{center}
\epsfxsize=3.0in\hspace{0in}\epsffile{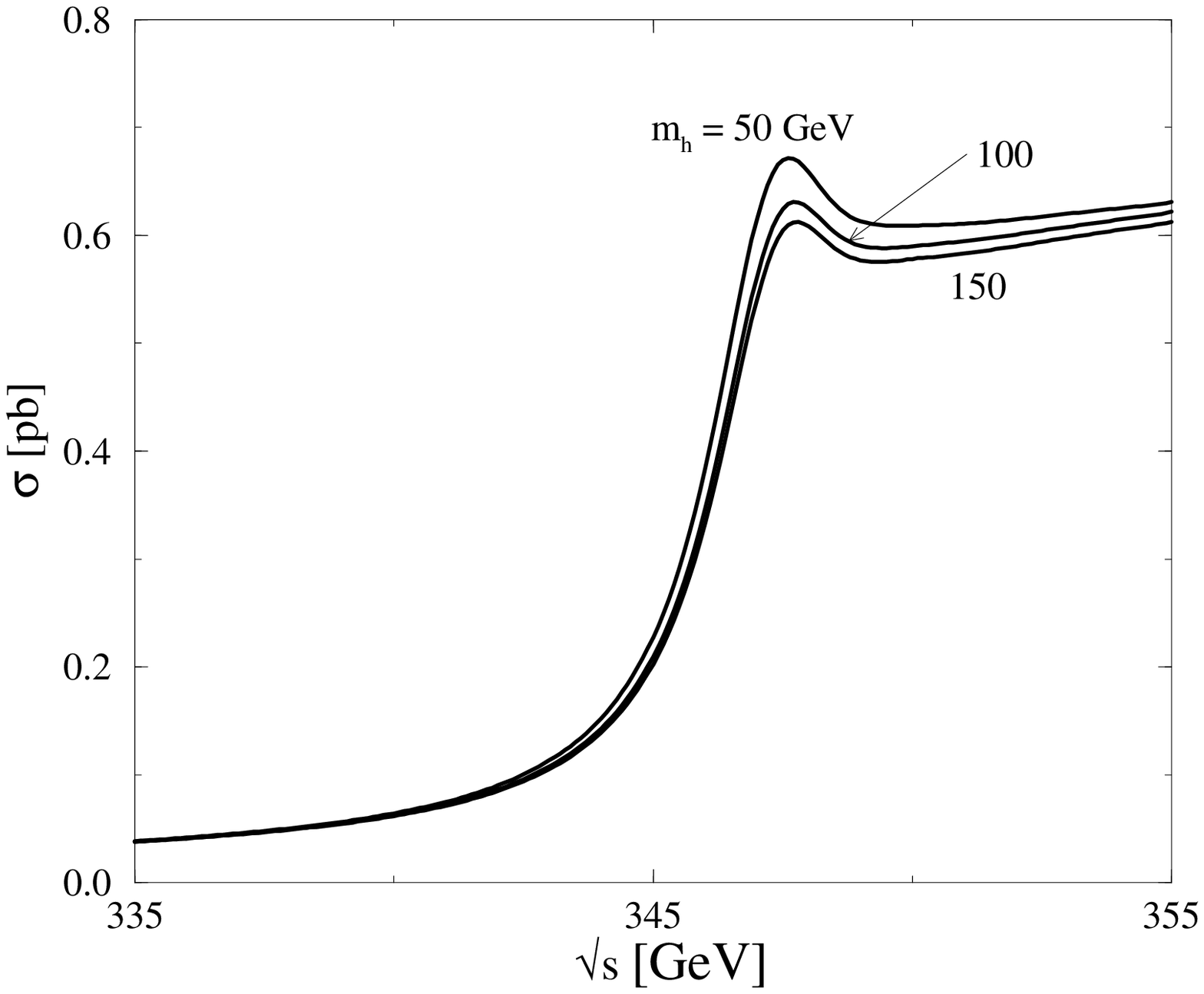}
\end{center}
\caption[]{\footnotesize\sf The dependence of the threshold region
on the Higgs mass, for $m_h=50, 100, 150$~GeV. Effects of
ISR and beam smearing have been included,
and we have assumed $m_t=175$~GeV and $\alpha_s(M_Z)=0.12$.}
\label{figure8}
\end{figure}

Since the exchange of a light Higgs boson can affect the threshold shape,
a scan of the
threshold cross section can in principle yield some information about the Higgs
mass and its Yukawa coupling to the top quark.
Figure~\ref{figure8} 
shows the dependence of the threshold curve on the Higgs mass, $m_h$.
However, it may be difficult to disentangle such a Higgs
effect from two-loop QCD effects, which are not yet fully
calculated~\cite{hoang}. 

QCD measurements at future colliders and
lattice calculations will presumably determine
$\alpha_s(M_Z)$ to 1\% accuracy (e.g.\ $\pm 0.001$) \cite{alphas}
by the time muon colliders are constructed so the uncertainty in
$\alpha_s$ will likely be similar to
the precision obtainable at a $\mu^+\mu^-$ and/or $e^+e^-$
collider with 100~fb$^{-1}$ integrated luminosity.
If the luminosity available for the threshold measurement is
significantly less than 100~fb$^{-1}$,
one can regard the value of $\alpha_s(M_Z)$ coming from
other sources as an input, and thereby improve the
top-quark mass determination.

There is some theoretical ambiguity in the
mass definition of the top quark. The theoretical
uncertainty on the quark pole mass
due to QCD confinement effects is of  order
$\Lambda_{QCD}$, {\it i.e.},  a few hundred MeV \cite{sw}.
In the $\overline{\rm MS}$ scheme of quark mass definition, the theoretical
uncertainty is better controlled.

Systematic errors in experimental efficiencies are not a significant
problem for the $t\anti t$ threshold determination of $\mt$.  This can
be seen from Fig.~\ref{figure6}, 
which shows that a 200 MeV shift in $\mt$ corresponds
to nearly a 10\% shift in the cross section
on the steeply rising part of the
threshold scan, whereas it results in almost no change in $\sigma$
once $\rts$ is above the peak by a few GeV.  Not only will efficiencies
be known to much better than 10\%, but also systematic uncertainties
will cancel to a high level of accuracy in the ratio of
the cross section measured above the peak to measurements
on the steeply-rising part of the threshold curve.

As Fig.~\ref{figure8} shows,
it will be important to know the Higgs mass and the $h t\anti t$
coupling strength in order to eliminate this source of systematic
uncertainty when extracting other quantities.

The measurements described in this section can be performed at either an
$e^+e^-$ or a $\mu^+\mu^-$ collider.
The errors for $\mt$ that we have found for the muon collider are smaller than
those previously obtained in studies at the NLC electron
collider primarily because
the smearing of the threshold region by the energy spread
of the beam is much less, and secondarily due to the fact that the reduced
amount of initial state radiation makes the cross section somewhat larger.

\section*{Conclusion}

With an integrated luminosity of 10 (100)~${\rm fb}^{-1}$,
the top-quark mass can be measured to 200 (70)~MeV,
using a 10-point scan over the threshold region, in 1~GeV intervals,
to measure the shape predicted by the QCD potential.
In the $t \bar{t} $ threshold study, differences of cross sections at energies
below, at, and above the resonance peak, along with the location of the
resonance peak, have different dependencies on the parameters
$m_t$, $\alpha_s$, $m_h$ and $|V_{tb}|^2$
and should allow their determination.
To utilize the highest precision measurements achievable at the statistical
level, theoretical uncertainties
and other systematics need to be under control. We are confident
that uncertainty in $\alpha_s$ will not be a factor and we have
noted that ratios of above-peak
measurements to measurements on the steeply rising part of the
threshold cross section will eliminate many experimental systematics
related to uncertainties in efficiencies.

\section*{Acknowledgments}

I thank V.~Barger, J.~F.~Gunion and T.~Han for a pleasant collaboration
on the issues reported here.
This work was supported in part by the U.S. Department of Energy
under Grant
No.~DE-FG02-91ER40661.

\end{document}